\documentclass[12pt,preprint]{aastex}

\newcommand{\refe}{} 
\newcommand{\refee}{ } 

\shorttitle{Observational Evidence Supporting the AEI}
\shortauthors{Mikles et al.}

\begin{document}

\title{QPO Frequency - Color Radius Connection in GRS~1915+105: a Possible Turnover supporting AEI predictions}

\author{Valerie J. Mikles}
\affil{Department of Astronomy, University of Florida, Gainesville, FL 32611}
\email{mikles@astro.ufl.edu}

\and 

\author{Peggy Varniere}
\affil{AstroParticule \& Cosmologie (APC), UMR 7164, 
Universit\'e Paris Diderot, 10 rue Alice Domon et L«eonie Duquet, 75205 Paris Cedex 13, France}
\email{varniere@apc.univ-paris7.fr}

\and 
\author{Stephen S. Eikenberry}
\affil{Department of Astronomy, University of Florida, Gainesville, FL 32611}
\email{eiken@astro.ufl.edu}

\and 

\author{Jerome Rodriguez}
\affil{CEA Saclay, DSM/IRFU/SAp, Laboratoire AIM
F-91191 Gif sur Yvette France}
\email{jerome.rodriguez@cea.fr}

\and 

\author{Dave Rothstein}
\affil{Department of Astronomy, Cornell University, Ithaca, NY 14853}
\email{droth@astro.cornell.edu}

\begin{abstract}

It is widely believed that the low frequency quasi-periodic X-ray oscillations observed in microquasars are correlated to, 
but  do not originate at, the physical radius of the inner edge of the accretion disk. Models relating the QPO frequency 
and color radius are hindered by observations showing contradicting trend correlations between the microquasars
 GRO 1655-40, XTE J1550-564 and GRS~1915+105. The first shows a negative correlation and the latter two a positive 
 one. By taking into account relativistic rotation in the accretion disk, the Accretion-Ejection Instability (AEI) model 
 predicts a turnover in the frequency-radius relationship, and has been successfully compared
 with observations of GRO J1655-40 and GRS 1915+105. We present further evidence supporting the AEI model 
 prediction by using observations of the microquasar GRS 1915+105. By combining a data set including 
$\theta$-,  $\beta$- and $\alpha$-class X-ray light curves, we observe positive, negative and null correlations in the 
 frequency-radius relationship. This is the first time a single source has shown a possible inversion in the QPO frequency-color 
 radius curve predicted by the AEI model.

\end{abstract}


\section{Introduction}

Microquasars are  X-ray binary stars featuring both accretion onto a compact object and relativistic jet ejection. 
To date, only a handful of microquasars are known. The most extensively studied is the archetype microquasar GRS 1915+105, 
which was discovered in 1992 \citep{CT92} and has been since observed at multiple wavelengths from the radio to the gamma-ray. 
Continued radio, infrared, and X-ray observations have led to the classification of GRS~1915+105's radio jets into three types: 
(1) ``steady'' radio jets, (2) discrete plasma ejection events of $20-40$ minute duration in the infrared and radio, 
and (3) large superluminal jets akin to the 1994 event that earned GRS~1915+105 the name ``microquasar'' \citep{mirabel94}. 
The discrete jets are associated with radio, infrared, and X-ray oscillatory behavior. 
The discrete jets are compact (extending a few hundred AU), and have a velocity $\sim$90\% of the speed of light. 
They range in strength from $5-200$~mJy in the infrared and radio \citep[e.g.,][]{eiken98,droth,yadav06}.

It is well established that the infrared and radio flares marking the discrete jet ejections in the source are associated 
with spectrally-hard dips in the X-ray light curves \citep{eiken98, mirabel98, fender98, kleinwolt02,rodriguez08}. 
These spectrally-hard dips are also associated with $2-10$~Hz variable low frequency quasi-periodic oscillations (LFQPOs). 
Despite the number of multi-wavelength observations of discrete jet ejection events in GRS~1915+105, the 
source of the LFQPO and the mechanism for launching the relativistic jet remain mostly unknown. 

Models explaining the LFQPO tend to tie the QPO frequency to a magnetoacoustical frequency \citep[e.g.,][]{titar04} 
or to the Keplerian frequency at some radius in the disk \citep[e.g.,][]{tagger99}. 
{\refe In the case of neutron star a different approach to the low frequency modulation was proposed, where the low frequency, normal branch oscillation, is
 a consequence of non-linear interaction between the KHz QPOs \citep{H04}.}
Previous observations have shown that the color radius of GRS~1915+105 is consistent with a monotonic Keplerian scaling. 
However, observations of the microquasar GRO 1655-40 show an inverted relationship compared with other
microquasars such as XTE J1550-564 or  GRS~1915+105  \citep{S00}.
In 1999, Tagger \& Pellat proposed the Accretion-Ejection Instability (AEI) as a possible explanation for the LFQPOs in 
microquasars. {\refe It was later found that the AEI predict an inversion in the QPO frequency - color radius relationship \citep{varniere02}
which was observed during the 1998 outburst of GRO J$1655$-$40$.}
In this letter we present new evidence that the turnover predicted by the AEI model is observed in the source GRS 1915+105.
 Our observational data set covering multiple X-ray states shows the turnover. This is the {\it first} recorded inversion in the 
 QPO frequency-inner radius trend found within a single source and lends credence to the AEI model

\section{The AEI Model}

The overall AEI model is described extensively in \citet{tagger99} and reviewed briefly in \citet{VT07}. 
\citet{CT01} perform 2D numerical  simulations of the AEI. 
\citet{T04} presented a scenario for the $\beta$ class of  GRS~1915+105 based on the identification of the LFQPO with the AEI.
Here we will summarize the salient points and the  observational tests of the AEI as the origin of the LFQPO. 

 The AEI is a global instability occurring in the inner region of a magnetized disk close to the equipartition 
namely when the magnetic pressure is of the order of the gas pressure. 
It is characterized by a spiral wave developing in the inner region of the disk. 
At the corotation radius between the accreting gas and the spiral wave, a Rossby vortex develops and stores accretion 
energy and angular momentum. In the presence of a low density corona the Rossby vortex will twist the
foot-point of the magnetic field line. This causes an Alfven wave to be emitted toward the corona, therefore linking accretion and ejection\citep{VT02}.

This instability is stronger when the magnetic field  is of the order of  equipartition with the gas pressure and require       
\begin{eqnarray}
\frac{\partial}{\partial \ln r} \ln \left(\frac{\kappa^2 \Sigma}{2\Omega B^2}\right) > 0
\end{eqnarray}
where  $\kappa$ and $\Omega$ are is the epicyclic and rotation frequency (in a Keplerian 
disk $\Omega = \kappa$), $\Sigma$ is the surface density and $B$ is the equilibrium magnetic field. 
{\refe  This configuration and  requirement of equipartition, although unusual for disk where
turbulence is studied,  are coherent with most of the MHD jet model and therefore adapted to the 
state we are studying. Also, because of the profile of magnetic field, the equipartition only happen in 
the inner most region of the disk leaving the outer part of the disk with a low (sub-equipartition) 
magnetic field. }

The AEI is a promising explanation for the LFQPO. First of all it is an  instability, therefore it   
does not require external excitation, but grows naturally.
It is also able to account for the following observational characteristics:
\begin{description}
   \item[\tt -] the rotation frequency of the dominant $m=1$ mode, {\em i.e} the one-armed spiral, 
    		predicted by the AEI is a few tenths of the Keplerian frequency at the inner edge of the disk. 
    		This frequency is consistent with the LFQPO frequency \citep{tagger99}.
   \item[\tt -]  based on variations in disk properties at the location of the spiral wave, 
             	the AEI partially reproduces the observed X-ray flux modulation. 
   \item[\tt -] non-linear simulations show that the rotating pattern of the AEI remains nearly steady, and thus is able to account for 
              persistent LFQPOs \citep{CT01}
   \item[\tt -] the AEI transfers energy and angular momentum toward the corona by Alfven Waves, thus providing a 
              supply of Poynting flux that may produce the compact jet often observed in the low-hard state  \citep{VT02}.
   \item[\tt -] by including General Relativity through the existence of a last stable orbit and orbital velocity profile, the AEI explains the 
   		observed turnover in the correlation between the color radius 
                   (= inner disk radius, as determined by the spectral fits) and the LFQPO frequency \citep{rodriguez02,varniere02}.
\end{description}

Because of these promising features, we are motivated to further study the AEI as a LFQPO model and 
compare the predictions to observations. In particular, we are interested in finding objects which show all of the aspects of 
the QPO frequency - inner radius relationship predicted by the AEI.

\section{GRS 1915+105 X-ray Observations}

GRS~1915+105 reveals its complexity through a broad display of multi-wavelength behavior. For a more complete review, see  
\citet{muno99,bell00,eiken00,droth}. To summarize, GRS~1915+105 has wild X-ray variability and several distinct X-ray 
light-curve classes with unique (but repeatable) appearance, count rate, and color. Although there are at least 12 distinct light-curve 
classes \citep{bell00}, our analysis focuses on three which have been positively linked to the discrete 
jet ejections: $\beta$-, $\alpha$-, and $\theta$-class. These three classes have spectrally hard dips followed by soft X-ray 
flares and infrared flares. For more complete reviews of the associated X-ray and infrared behavior, see \citet{eiken00,droth,mikles06}.

The X-ray data analysis is described in detail in \citet{mikles06}, and summarized here. From three RXTE observations 
of GRS~1915+105 taken on 14 August 1997, 9 September 1997, and 10 July 2002\footnote{This observation is composed of several
small window. We only too the beginning of the observation which was in the $\theta$-class.}, listed in Table \ref{tbl-1}, we  extract 
Proportional Counter Array (PCA) Standard-1 light curves and identify regions showing a hard X-ray dip. The regions were 
chosen as representative of the X-ray behavior associated with the particular light-curve state at the time of the observations: 
one $\beta$-class, one $\alpha$-class, and one $\theta$-class. In addition to the spectrally hard dip region for each class, we also examine the X-ray oscillations in the $\alpha$-class that precede the spectrally hard dip. This region shows brief hard dips associated with low-frequency QPOs and X-ray mini-flares \citep{droth}. For each observation, we extract binned mode 8-millisecond light 
curves in the $2 - 13$ keV range and 4-second resolution binned and event X-ray spectra in the $2 - 25$ keV range.  

By using a FFT, we calculate the power density spectrum (PDS) from the binned mode 8-millisecond light curve. We fine-bin the PDS using Fourier interpolation and track the peak frequency at 4-second resolution \citep[see e.g.][]{ransom02}. We determine the QPO~frequency by fitting a Moffat function to the PDS in the $2 - 10$~Hz frequency range. Further discernment of the QPO frequency from noise peaks for the $\alpha$- and $\beta$-class hard dip observations is described in \citet{mikles06}. For the $\alpha$-class oscillation region and the $\theta$-class dips, the QPO is clearly present above 4~Hz, and lower frequency peaks are discounted as noise.

\begin{figure*}
\includegraphics[width=5in]{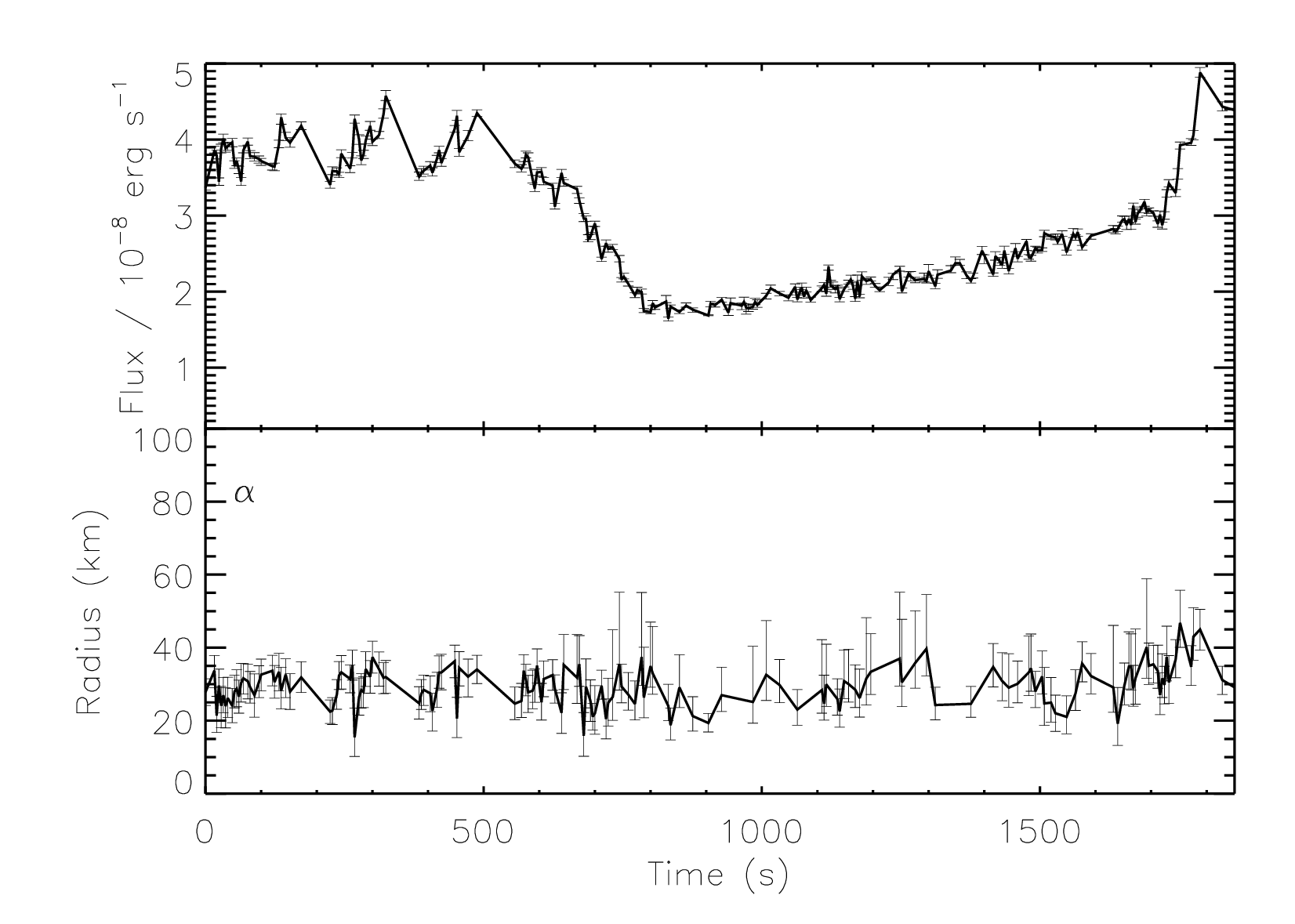}
\includegraphics[width=5in]{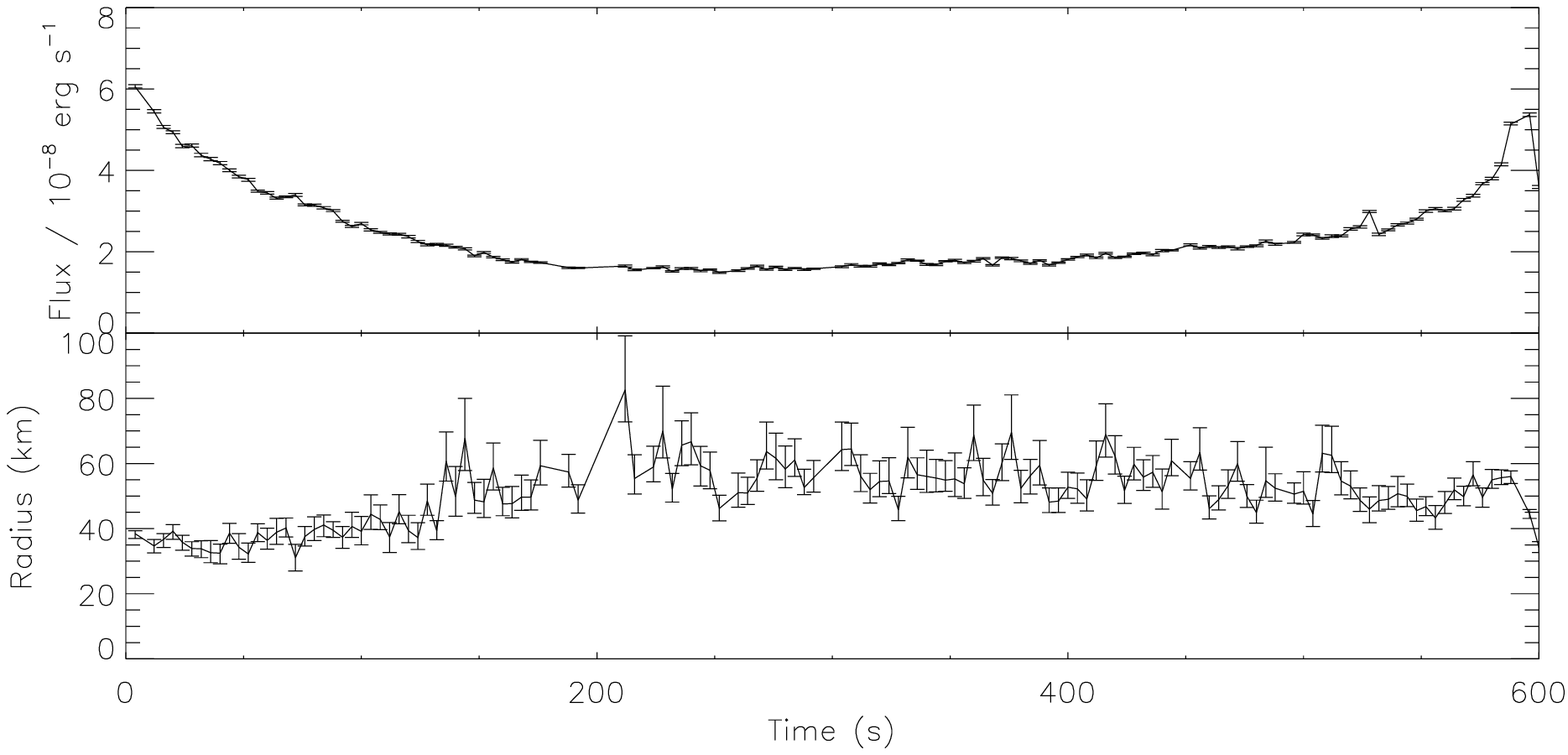}
\includegraphics[width=5in]{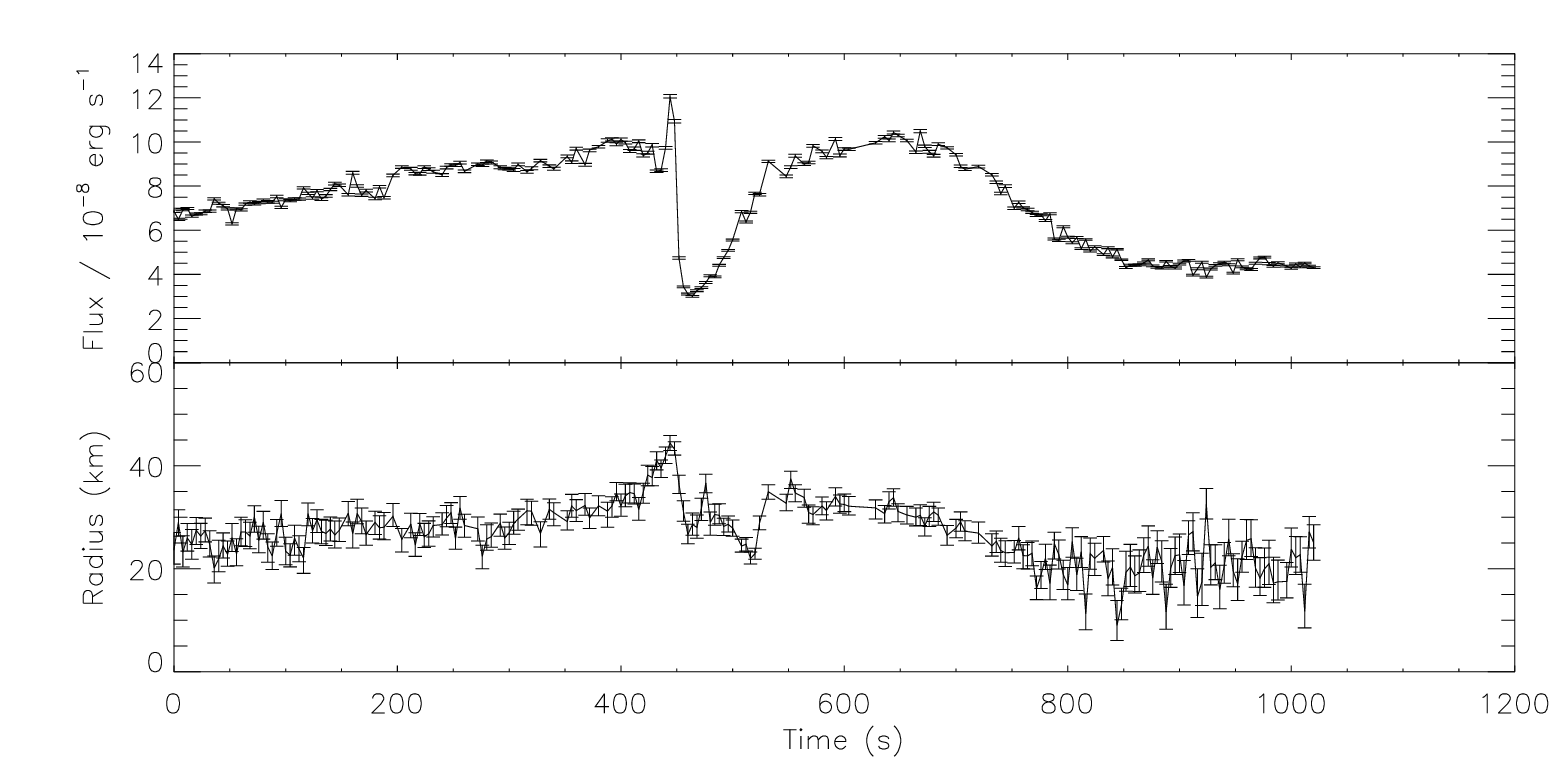}
\caption{Light curve and radius evolution for $\beta$-, $\alpha$-, and $\theta$-class light curves.}
\label{fig:rad}
\end{figure*}

{\refee  We fit the spectra with a combination of absorbed multi-temperature disk blackbody and power law models (\texttt{wabs*(diskbb+powerlaw)}).
 The multi-temperature blackbody models the thermal emission from a geometrically thin accretion disk around a compact object, recognizing
that the temperature of the disk increases at low radii close to the compact object. This model is characterized by the X-ray flux, the temperature and and
radius at the inner part of the accretion disk. We exclude spectral fits with $\chi ^2 > 2$ or with poorly constrained
blackbody radii, and are left with over 200 data points to build the relation between the frequency of the QPO and the inner radius of the
accretion disk, shown in Fig. \ref{fig:rad}.

The choice of the simple black body model used in fitting the spectra is dictated by two things. First, rather than
testing different models we simply want to compare the results from different classes, and therefore
we chose to use the simplest, standard model of \texttt{diskbb} from \texttt{XSPEC}. Second, since our
spectra are taken at 4s intervals, we would gain relatively little by employing more complicated
models. Although this model oversimplifies the complex dynamics of the magnetized accretion disk, it allows for greater
consistency in the analysis with a broader sample and eliminates the need to deconstruct partial
correlation of features in a more complex model. }

In Figure \ref{fig:rad}, we plot the X-ray light-curve and radius evolution for the 
three data sets used in this analysis. 

The color radius is determined by fitting a geometrically thin accretion disk model which, 
while fairly standard, oversimplifies  the complex dynamics of the magnetized accretion disk. Still, 
when we fit the X-ray spectra using XSPEC 11.3, the blackbody normalization gives a radius often too small.
 \cite{merloni00} performed a reliability study of the radii determination using {\tt diskbb}  in XSPEC. 
 They created a series of model X-ray spectra,  added noise and accounted for detection quirks, and found that 
 the XSPEC radius fits were  not fully reliable under a low disk flux condition. 
 They also found that no single 
 correction factor can be applied in that case to correct the fitted color radius to the inner radius. 
 Nevertheless, the {\tt diskbb} model often give acceptable value for the radius (within was could be expected
 knowing the mass), even in the Hard state where the disk flux is low{\refe , and allow simpler  comparison between different
 observations}. Also, in the Very High state, or Steep Power Law
 (state with both the disk and the power-law contribute significantly to the spectrum) the {\tt diskbb} model
 systematically underestimates the disk radius. 
  
So long as the XSPEC model assumptions are reasonable for the system, we expect some correlation to exist 
between the color radius, $R_{col}$, and the inner disk radius, $R_{in}$. Thus, even though the value of the 
color radius is low, it still traces the evolution of the inner disk. Often, the color radius and inner disk radius are 
related by a hardening factor, $f$, such that $R_{in} = R_{col} \times f^2$ \citep{st95}. For a more complete 
discussion of the hardening factor, see, e.g., \citet{st95, merloni00, rodriguez02}. The value of the hardening factor 
is found to be $f \sim 1.7-2.0$, and depends on several physical parameters in the system, such as viscosity and
accretion rate \citep{st95}. Given that we are measuring the QPO frequency - radius relationship while in a 
spectrally hard dip, where the blackbody emission contributes weakly to the total flux, our radius fits should be 
taken with caution. However, we proceed with our analysis assuming that $R_{col}$ traces the trend of the 
inner radius evolution with reasonable accuracy.

\section{Radius-Frequency plot: data and theory}

\begin{figure*}[!hb]
\includegraphics[width=6in]{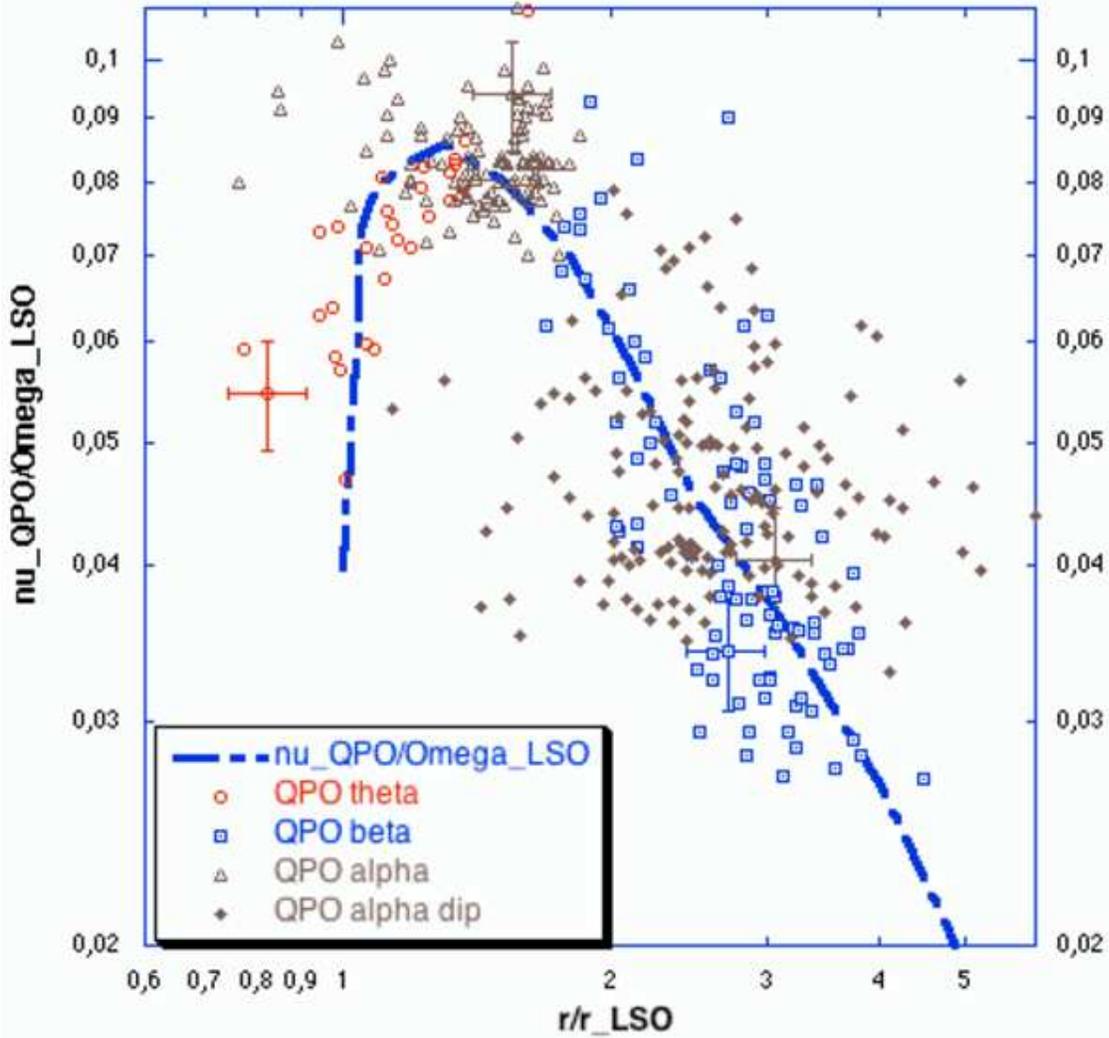}
\caption{Radius (in $r_{LSO}$) vs. QPO frequency (in frequency at the last stable orbit) on a log-log scale. The dip of the $\beta$ and $\alpha$-class data ccupies the Keplerian regime of the model. The oscillation at the beginning of the $\alpha$-class data occurs near the turnover. The $\theta$-class data is located where the trend is inverted}
\label{fig:theo}
\end{figure*}

In Figure \ref{fig:theo}, we show the predictions of the AEI against the observational data presented above. 
We see that, at higher color radii $R_{col} > 3 r_{LSO}$  (about  60~km in the case of GRS 1915+105), the model predicts a Keplerian relationship $\nu_{QPO} \propto R_{col}^{-3/2}$. For  small radii  the AEI prediction deviates from a purely Keplerian relationship. At radii between   $1.3 r_{LSO} < R_{col} < 3 r_{LSO}$  (about  40~km$ < R_{col} <$ 60~km) the  
deviation is small, and for  lower radii ($R_{col} <$ 40~km), the correlation reverses.   

The three classes are ordered around the curve. The $\theta$-class observations are on the left, 
meaning that in this state, the inner edge of the accretion disk extends  closer to its last stable orbit. 
The oscillation at the beginning of the $\alpha$-class observations cluster toward the top of the curve, at a relatively constant QPO frequency
while the later dip is better fitted on the keplerian side.
Finally, the $\beta$-class observations fall in the more usual, close to Keplerian, 
 portion of the curve as was observed before \citep{rodriguez02}. 
The discovery of this inversion in the QPO frequency-inner radius trend is the {\it first} record of an inversion within a 
single source and lends credence to the AEI model. The AEI model is currently one of the only QPO frequency models that 
accounts for this heretofore unexplained reversal in the frequency-radius trend.


Although there is an apparent spread of the observational points around the model predictions, it is a serious 
concern only at low radii where the fits are known to have difficulties.  We kept all the points, even the one with an unrealistically  
low color radii compared with the Schwarzshild radii expected from a $14\ M_\odot$ black hole.

Still, the AEI model is an improvement over LFQPO models that predict simple Keplerian scaling in the 
LFQPO frequency - radius relationship. Despite inherent problems in radius estimation, it is interesting 
(and promising) that the $\theta$-class data, which have lower radii than the $\beta$-class data, fall within the 
same 2 - 10 Hz frequency range and are hence consistent with the turnover predicted in the AEI model. 
It suggests that while the instantaneous estimates of the color radius may be in error, the {\it trend} predicted by 
the XSPEC model fits is statistically accurate and useful for testing models. This new evidence of a turnover in the QPO frequency - color radius trend is a very exciting step toward validating the AEI model.

Notably, in the $\theta$-class, the inner radius always stays close to the last stable orbit which is often
referred as the position where the  High-Frequency QPO form \citep[see e.g.][]{TV06}.  
Recent observations by \citep{B06} confirmed the presence of  HFQPO in the $\theta$-class.
Following \citet{TV06},  HFQPOs can occur when the inner edge of the disk is less than 
$1.3\ r_{LSO}$ which corresponds to the maximum of $\kappa$, but with a lower amplitude as the inner edge  of the disk 
gets away from the  last stable orbit.

\section{Conclusions}

We observe three different classes of X-ray oscillation in GRS~1915+105 and find that our QPO~frequency - color radius correlations fall over multiple regimes of the theoretical curve predicted by the AEI model. Although the {\it trend} we observe in the QPO~frequency - radius relation of GRS~1915+105 should be taken with caution, especially at low radii, this set of observations captures a very important observational prediction of the AEI model and behaviorally links 
GRS~1915+105 to the microquasar GRO~1655-40. We can further refine this result by gathering archival data of 
GRS~1915+105 in different X-ray spectral states, specifically targeting times when multi-wavelength observations are 
available. In this way, we can test the validity of the model in predicting both jet ejection strength and the QPO behavior. 
Additionally, we can expand the analysis to multiple microquasar systems and test the applicability of the model. 
For example, the microquasar GRO~1655-40 has been observed in the non-Keplerian regime, but additional observations 
may show that this source, like GRS~1915+105, can display behaviors consistent with multiple regimes. In comparing 
multiple sources exhibiting QPOs, we can determine whether most X-ray binary sources stay confined to a single part of 
the frequency-radius curve or whether all microquasars have a range of variability (and associated jet ejections) similar to 
GRS~1915+105. Ultimately, by testing observations against theory, we can determine whether we are able to accurately predict 
larger jet ejections via QPO observations so that we can study ejection events more closely.

\acknowledgements
VJM and SSE are supported in part by NSF grant AST-0507547.

\clearpage

\begin{center}
\begin{table}
\begin{tabular}{lllll}
\bf{Class} & \bf{RXTE DATA ID} & \bf{Date Observed} & \bf{Start Time} \\
\hline
\hline
$\beta$   & 20186-03-03-01 & 1997 Aug 14        & 04:20:52  \\
\hline	     						      
$\alpha$ & 50125-01-04-00 & 2002 Jul 27        & 07:15:00  \\
\hline	     						      
$\theta$ & 30182-01-03-00 & 1998 Jul 10        & 05:05:57  \\
\hline

\end{tabular}
\caption{Observation IDs and dates of the three data sets. The Start Time indicates where spectral fitting began. The class is based on the \cite{bell00} system.}
\label{tbl-1}
\end{table}
\end{center}

\clearpage

\end{document}